\documentclass[twocolumn,showpacs,showkeys,amsmath,amssymb]{revtex4}

\usepackage{graphicx}

\begin{document}

\title{ Quasi-Equatorial Gravitational Lensing by Spinning Black Holes
\\ in the Strong Field Limit}

\author{V. Bozza}

\email{valboz@sa.infn.it}

\affiliation{Centro Studi e Ricerche ``Enrico Fermi'', Rome,
Italy.\\ Dipartimento di Fisica ``E.R. Caianiello'', Universit\`a
di Salerno, Italy.\\
   Istituto Nazionale di Fisica Nucleare, Sezione di
 Napoli.}

\date{\today}

\begin{abstract}
Spherically symmetric black holes produce, by strong field
lensing, two infinite series of relativistic images, formed by
light rays winding around the black hole at distances comparable
to the gravitational radius. In this paper, we address the
relevance of the black hole spin for the strong field lensing
phenomenology, focusing on trajectories close to the equatorial
plane for simplicity. In this approximation, we derive a
two-dimensional lens equation and formulae for the position and
the magnification of the relativistic images in the strong field
limit. The most outstanding effect is the generation of a non
trivial caustic structure. Caustics drift away from the optical
axis and acquire finite extension. For a high enough black hole
spin, depending on the source extension, we can practically
observe only one image rather than two infinite series of
relativistic images. In this regime, additional non equatorial
images may play an important role in the phenomenology.
\end{abstract}



\maketitle

\section{Introduction}

Gravitational lensing, since its beginning, has been used to test
General Relativity in the weak field approximation. Later on,
several studies of null geodesics in strong gravitational fields
have been lead in past years. Bardeen et al. \cite{Bar} studied
the appearance of a black hole in front of a uniform source;
Viergutz \cite{Vie} made a semi-analytical investigation of null
geodesics in Kerr geometry; Nemiroff \cite{Nem} studied the visual
distortions around a neutron star and around a black hole; Falcke,
Melia \& Agol \cite{FMA} considered the luminosity of the
accretion flow as a source.

A recent paper by Virbhadra \& Ellis \cite{VirEll} has risen a new
interest about gravitational lensing as a probe for strong
gravitational fields generated by collapsed objects, providing a
new important test for the full general relativity. They have
shown that a source behind a Schwarzschild black hole would
generate an infinite series of images on both sides of the lens.
These relativistic images are formed by light rays passing close
to the event horizon and winding several times around the black
hole before emerging towards the observer. By an alternative
formulation, Frittelli, Kling \& Newman \cite{FKN} attained an
exact lens equation, giving integral expressions for its
solutions, and compared their results to those by Virbhadra \&
Ellis. The phenomenology of collapsed objects with naked
singularities, analyzed by Virbhadra \& Ellis in another work
\cite{VirEll2}, is radically different. This difference provides a
possible way to test the correctness of the cosmic censorship
conjecture.

A new, simple and reliable method to investigate the subject was
proposed by Bozza et al. in Ref. \cite{BCIS}. They revisited the
Schwarzschild black hole lensing defining a {\it strong field
limit} for the deflection angle, which retained the first two
leading order terms. By this approximation, a fully analytical
treatment was developed and simple formulae for the position and
the magnification of the images were derived. The same method was
applied by Eiroa, Romero \& Torres \cite{ERT} to a
Reissner-Nordstrom black hole, confirming the appearance of a
similar pattern of images. Later on, the formulae given in Ref.
\cite{BCIS} were used by Petters \cite{Pet} to calculate
relativistic effects on microlensing events. Finally, in a
previous work, we have developed a generalization of the strong
field limit to an arbitrary spherically symmetric spacetime
\cite{Boz}, comparing the image patterns for several interesting
metrics. We have shown that different collapsed objects are
distinguishable by a careful examination of the separation between
the first two relativistic images and their luminosity ratio. The
sensitivities required for such measures are out of the actual
VLBI projects \cite{VLBI}, but might be reached in a not so far
future.

Insofar, only spherically symmetric black holes have been
adequately investigated. Yet, in general, a black hole would be
characterized by a non-zero intrinsic angular momentum which
breaks spherical symmetry, leaving only a rotational symmetry
around one axis. Rotation heavily affects the gravitational field
around the collapsed object. It is thus natural to expect relevant
modifications in the phenomenology of strong field gravitational
lensing.

A further motivation for such a study comes from the fact that
previous works on the subject have selected the supermassive black
hole hosted by the radio source Sagittarius A* \cite{Ric} as the
best candidate for strong field gravitational lensing
\cite{VirEll,Boz}. Its mass has been estimated to be $M=2.6 \times
10^6 M_\odot$ but our knowledge on a possible intrinsic angular
momentum of this object is still very poor. However, the analysis
of the variability of the spectrum in the mm-submm region suggests
the possibility of a non-negligible spin \cite{LiuMel}. In
particular, the value
\begin{equation}
|a|\approx 0.044
\end{equation}
has been proposed (with respect to Ref. \cite{LiuMel}, we normalize
distances to the Schwarzschild radius rather than to the black
hole mass, hence the factor $1/2$ in this value), but high
uncertainties in the assumptions behind the calculations may push
the spin towards even higher values.

The purpose of this paper is to investigate the relevance of the
black hole spin in strong field lensing phenomenology. We
formulate the strong field limit for Kerr black hole lensing,
considering trajectories close to the equatorial plane. With this
limitation we cannot give a complete description of the whole
phenomenology, which we shall delay to future works. However, as
we shall see, quasi-equatorial motion yields very sharp
indications useful to understand also the general case. In
particular, we will show that the presence of a sufficiently high
black hole spin drastically changes the expected pattern of
observable images.

This paper is structured as follows. In Sect. 2 we recall some
general properties of Kerr geodesics. In Sect. 3 we carry out the
strong field limit expansion of the deflection angle on the
equatorial plane. In Sect. 4 we move off from the equatorial plane
and consider trajectories at small declinations. In Sect. 5 we
write the lens equation on the equatorial plane. In Sect. 6 we
write a polar lens equation dealing with displacements normal to
the equatorial plane. In Sect. 7 we find the positions of the
caustic points and the magnification of all images in the
equatorial plane. In Sect. 8 we describe the critical curves and
caustic structure. Sect. 9 discusses the effects of the black hole
spin on the gravitational lensing phenomenology, with special
reference to the black hole at the center of our Galaxy. Sect. 10
contains the summary.

\section{Geodesics in Kerr spacetime}

In Boyer-Lindquist coordinates \cite{BoyLin} $x^\mu \equiv
(t,x,\vartheta,\phi)$, the Kerr metric reads

\begin{eqnarray}
&& \! \! \! \! \! \! \! \! \! \! \! \! ds^2=\frac{\Delta-a^2
\sin^2 \vartheta}{\rho^2}d t^2-\frac{\rho^2}{\Delta} dx^2-\rho^2
d\vartheta^2- \nonumber \\ && \! \! \! \! \! \! \! \! \! \! \! \!
\frac{ \left(x^2+a^2 \right)^2 - a^2\Delta \sin^2 \vartheta
}{\rho^2} \sin^2 \vartheta d\phi^2+\frac{2ax
\sin^2\vartheta}{\rho^2} dt d\phi \\%
&& \! \! \! \! \! \! \! \! \! \! \! \!\Delta=x^2-x+a^2 \\%
&& \! \! \! \! \! \! \! \! \! \! \! \! \rho^2=x^2+a^2
\cos^2\vartheta
\end{eqnarray}
where $a$ is the specific angular momentum of the black hole. All
distances are measured in Schwarzschild radii
($\frac{2MG}{c^2}=1$).

The Kerr space is characterized by a spherical event horizon at
$x_H=\frac{1+\sqrt{1-4a^2}}{2}$ for $|a|<0.5$. Beyond this
critical value of the spin there is no event horizon and causality
violations are present in the whole spacetime, with the appearance
of a naked singularity \cite{Car,HawEll}. We shall restrict to
subcritical angular momenta. The ellipsoid $\Delta-a^2 \sin^2
\vartheta=0$ is the static limit bounding the region where every
static worldine ($x=\vartheta=\phi=0$) becomes spacelike. The
region between the static limit and the horizon is called
ergosphere: here everything is bound to rotate around the black
hole.

The geodesics equations can be derived taking
\begin{equation}
{\cal L}=g_{\mu\nu} \dot x^\mu \dot x^\nu
\end{equation}
as the Lagrangian, where the dot indicates the derivative with
respect to some affine parameter.

Finding four integrals of motion, we can transform these equations
into a set of four first order equations which are equivalent to
the original ones. Two costants of motions are the energy and the
angular momentum of the particle, given by
\begin{eqnarray}
&& 2E= \frac{\partial {\cal L}}{\partial t}  \label{E}\\ %
&& -2J= \frac{\partial {\cal L}}{\partial \phi}. \label{J}
\end{eqnarray}
By a suitable choice of the affine parameter, we can set
\begin{equation}
E=1.
\end{equation}

From these equations, we find an expression for $\dot t$ and $\dot
\phi$ in terms of $x$, $\vartheta$ and $J$
\begin{eqnarray}
&& \dot t= \frac{g_{33}+g_{03} J
}{g_{33}g_{00}-g_{03}^2} \label{dtds} \\%
&& \dot \phi= \frac{g_{00} J+g_{03} }{g_{03}^2-g_{33}g_{tt}}.
\label{dphids}
\end{eqnarray}

${\cal L}$ is another constant of motion, which vanishes for null
geodesics and can be used to write $\dot x$
\begin{equation}
\dot x=\pm \sqrt{\frac{-g_{00}\dot t^2 -g_{22}\dot
\vartheta^2-g_{33}\dot \phi^2-2g_{03}\dot t \dot \phi}{g_{11}}}.
\label{dxds}
\end{equation}

Finally, $\dot \vartheta$ can be obtained in terms of a fourth
integral of motion, separating the Hamilton-Jacobi equation
\cite{Car}:
\begin{equation}
 \dot  \vartheta=\pm \frac{1}{\rho^2}\sqrt{Q+a^2\cos^2 \vartheta-J^2
\cot^2 \vartheta }. \label{dottheta}
\end{equation}

Eqs. (\ref{dtds})-(\ref{dottheta}) represent the sought set of
first order differential equations, suitable for a detailed study.

\begin{figure}
\resizebox{\hsize}{!}{\includegraphics{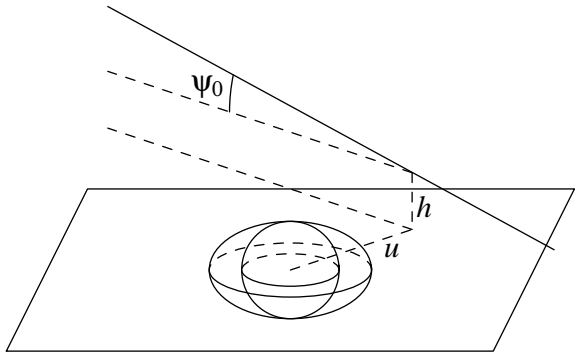}}
 \caption{The three parameters identifying an incoming light ray.
 }
 \label{Fig IniCon}
\end{figure}

The integrals of motion $J$ and $Q$ can be expressed in terms of
the geometric parameters of the incoming light ray trajectory. In
general, we can identify a light ray coming from infinity by three
parameters (Fig. \ref{Fig IniCon}), referring to the straight line
which the photon would follow if there were no gravitational
field. The projection of this line on the equatorial plane has a
distance $u$ from the origin, which we shall call the projected
impact parameter. At this minimum projected distance, the light
ray has some height $h$ on the equatorial plane. Finally, the
inclination $\psi_0$ is the angle that the light ray forms with
the equatorial plane. When we switch on the gravitational field,
the light ray is obviously deviated from this ideal straight line,
but these three parameters can still be used to label any light
rays coming from infinity.

Taking the asymptotic limit of the equations of motion, we can
write $J$ and $Q$ in terms of the initial conditions
characterizing the light ray
\begin{eqnarray}
&& J=u \cos{\psi_0} \label{J=u cos psi0} \\%
&& Q=h^2 \cos^2 \psi_0+(u^2-a^2) \sin^2 \psi_0. \label{Q}
\end{eqnarray}

\section{Deflection angle in the equatorial plane} \label{Sec equ}

In this section, we consider light rays strictly lying on the
equatorial plane $\vartheta=\frac{\pi}{2}$ by setting
$h=\psi_0=0$. The reduced metric has the form
\begin{equation}
ds^2=A(x)dt^2-B(x)dx^2-C(x)d\phi^2 +D(x) dt d\phi.
\end{equation}
with
\begin{equation}
\begin{array}{l}
  A(x) = 1 - \frac{1}{x}  \\ \\
  B(x) = \frac{1}{1-\frac{1}{x}+\frac{a^2}{x^2}} \\ \\
  C(x) = x^2+a^2+\frac{a^2}{x} \\ \\
  D(x) = 2\frac{a}{x}. \label{ABCD}
\end{array}
\end{equation}

What we say in this section is immediately extendable to any
axially symmetric spacetime if we replace Eq. (\ref{ABCD}) by any
other expression.

As $\psi_0=0$, by Eq. (\ref{J=u cos psi0}) the angular momentum
$J$ coincides with the impact parameter $u$. The impact parameter
is also related to the minimum distance $x_0$ reached by the
photon. In general, a light ray coming from infinity approaches
the black hole, reaches this minimum distance $x_0$ and then
leaves again towards infinity. Evaluating the Lagrangian at
$x=x_0$, we find an implicit relation between $J=u$ and the
closest approach distance $x_0$
\begin{eqnarray}
&J=u &=\frac{-D_0+\sqrt{4A_0 C_0 +D_0^2}}{2A_0}= \nonumber \\
&& =\frac{-a+x_0 \sqrt{a^2+x_0(x_0-1)}}{x_0-1},
\end{eqnarray}
where all the metric functions with the subscript 0 are evaluated at
$x=x_0$. The impact parameter $u$ is then univocally determined by
$x_0$ and vice versa. Choosing the positive sign before the square
root, we describe only light rays winding counterclockwise when
seen from above. For $a>0$ the black hole also rotates
counterclockwise, while for $a<0$ the black hole and the photon
rotate in opposite senses.

Dividing $\dot \phi$ by $\dot x$, we find the azimuthal shift as a
function of the distance
\begin{eqnarray}
&& \frac{d\phi}{dx}=P_1(x,x_0)P_2(x,x_0) \label{dphidx}\\%
&& P_1(x,x_0)=\frac{\sqrt{B}(2A_0 A J+ A_0 D)}{\sqrt{C A_0}
\sqrt{4AC+D^2}} \\%
&& P_2(x,x_0)=\frac{1}{\sqrt{A_0-A \frac{C_0}{C}+\frac{J}{C}(A
D_0- A_0 D )}}.
\end{eqnarray}

Integrating this expression from $x_0$ to infinity we find half
the deflection angle as a function of the closest approach. Given
the symmetry between approach and departure, we can write the
whole deflection angle as
\begin{eqnarray}
&& \alpha(x_0)=\phi_f(x_0)-\pi \\ %
&& \phi_f(x_0)=2\int\limits_{x_0}^\infty \frac{d\phi}{dx} dx.
\label{phif}
\end{eqnarray}

$\phi_f(x_0)$ is the total azimuthal shift. It evaluates to $\pi$
for a straight line and becomes larger as the light ray is bent by
the gravitational field. The expression for a spherically
symmetric metric, given in Ref. \cite{Boz} can be recovered
setting $D=D_0=0$.

The deflection angle grows as $x_0$ decreases. It diverges when
$x_0$ reaches a minimum value $x_m$ which represents the radius of
the photon sphere. If a photon falls inside this sphere, it is
destined to be absorbed by the black hole. Of course, we will have
different photon spheres for photons winding in the same sense of
the rotation of the black hole (hereafter direct photons)
and for photons winding in the opposite sense (retrograde photons). In
general, we expect the latter to be absorbed more easily. Their
photon sphere will thus be larger than that of retrograde
photons, which can escape more easily. As we shall see later, this
is what happens.

Following the philosophy of the strong field limit, we look for an
expansion of the deflection angle of the form
\begin{equation}
\alpha(\theta)=-\overline{a} \log \left(\frac{\theta
D_{OL}}{u_m}-1 \right) +\overline{b} +O\left(u-u_m \right)
\label{S F L theta}
\end{equation}
where the coefficients $u_m$, $\overline{a}$ and $\overline{b}$
depend on the metric functions evaluated at $x_m$. $D_{OL}$ is the
distance between the lens and the observer, so that the angular
separation of the image from the lens (also referred as impact
angle) is $\theta=\frac{u}{D_{OL}}$.

All the steps to be taken towards this final expression are very
similar to those for spherically symmetric black holes, with few
adjustments. We shall sketch them very briefly, referring to Ref.
\cite{Boz} for details.

We define the variables
\begin{eqnarray}
&& y=A(x) \\%
&& z= \frac{y-y_0}{1-y_0}
\end{eqnarray}
where $y_0=A_0$. The integral (\ref{phif}) in the deflection angle
becomes
\begin{eqnarray}
&& \phi_f(x_0)=\int\limits_0^1 R(z,x_0) f(z,x_0) dz \label{I z} \\%
&& R(z,x_0)=2\frac{1-y_0}{A'(x)}P_1(x,x_0) \label{R} \\%
&& f(z,x_0)=P_2(x,x_0) \label{f}
\end{eqnarray}
where $x=A^{-1} \left[\left(1-y_0 \right) z+ y_0 \right]$.

The function $R(z,x_0)$ is regular for all values of $z$ and
$x_0$, while $f(z,x_0)$ diverges for $z \rightarrow 0$. To find
out the order of divergence of the integrand, we expand the
argument of the square root in $f(z,x_0)$ to the second order in
$z$
\begin{equation}
 f(z,x_0) \sim f_0(z,x_0)= \frac{1}{\sqrt{\alpha z +\beta z^2}}.
 \label{f0}
\end{equation}

When $\alpha$ is non zero, the leading order of the divergence in
$f_0$ is $z^{-1/2}$, which can be integrated to give a finite
result. When $\alpha$ vanishes, the divergence is $z^{-1}$ which
makes the integral diverge. Then the outermost solution of the Eq.
$\alpha=0$ defines the radius of the photon sphere $x_m$ (see also
\cite{CVE}).

\begin{figure}
\resizebox{\hsize}{!}{\includegraphics{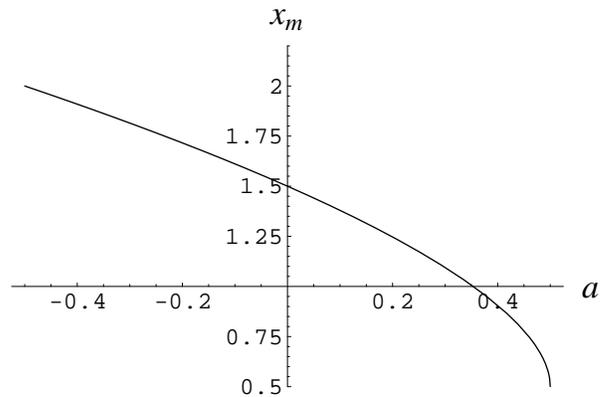}}
 \caption{The radius of the photon sphere versus the black hole angular momentum.}
 \label{Fig xm}
\end{figure}

In the case of the Kerr metric, we have
\begin{eqnarray}
&\alpha &=x_0 \left[x_0(3-5x_0+2x_0^2)-2a^2+
 2a
\sqrt{a^2+x_0(x_0-1)} \right] \cdot
\nonumber \\ &&\cdot \left[ (x_0-1)(x_0^3+a^2(x_0+1)
\right]^{-1}.
\end{eqnarray}
Eq. $\alpha=0$ is equivalent to the third degree equation
\begin{equation}
8a^2-x_0(3-2x_0)^2=0.
\end{equation}
The real solution external to the horizon of this equation defines the radius
of the photon sphere $x_m$, plotted in Fig. \ref{Fig xm}. As expected, for
positive $a$, (direct) photons are allowed
to get closer to the black hole, entering even the ergosphere
($x_m$ falls below 1) at high values of $a$. It is possible to
calculate exactly from Eq. $\alpha=0$ at what angular momentum
this happens. The critical value is
\begin{equation}
a_{cr}=\frac{1}{2\sqrt{2}}=0.354.
\end{equation}

Notice that $x_m \rightarrow 1/2$ as $a\rightarrow 1/2$, i.e. the photon
sphere coincides with the horizon in the limit of extremal Kerr black hole.
For negative angular momenta, (retrograde) light rays must keep
farther from the center.

The procedure to find the strong field limit coefficients is from
now on identical to that described in Ref. \cite{Boz}, with
$R(z,x_0)$ $f(z,x_0)$ and $f(z,x_0)$ given by Eqs. (\ref{R}),
(\ref{f}) and (\ref{f0}), respectively. We shall not repeat the
whole technique here but just specify the results of Ref.
\cite{Boz} for our metric.

The strong field limit coefficients of the
expansion (\ref{S F L theta}) are
\begin{eqnarray}
&& u_m=\frac{-D_m+\sqrt{4A_m C_m +D_m^2}}{2A_m} \\
&& \overline{a}= \frac{R(0,x_m)}{2\sqrt{\beta_m}} \label{ob1}\\%
&& \overline{b}=-\pi+b_D+b_R+\overline{a} \log \frac{c x_m^2}{u_m}
\label{ob2}
\end{eqnarray}
where
\begin{eqnarray}
&&\! \!\!\!\! b_D=2 \overline{a} \log
\frac{2(1-y_m)}{A'_m x_m}
\\ && \! \!\!\! b_R=\int\limits_0^1\left[
R(z,x_m)f(z,x_m)-R(0,x_m)f_0(z,x_m) \right]dz
\end{eqnarray}
and $c$ is defined by the expansion
\begin{equation}
u-u_m=c \left(x_0-x_m \right)^2. \label{xtou}
\end{equation}
All the functions with the subscript $m$ are evaluated at
$x_0=x_m$.

\begin{figure}
\resizebox{\hsize}{!}{\includegraphics{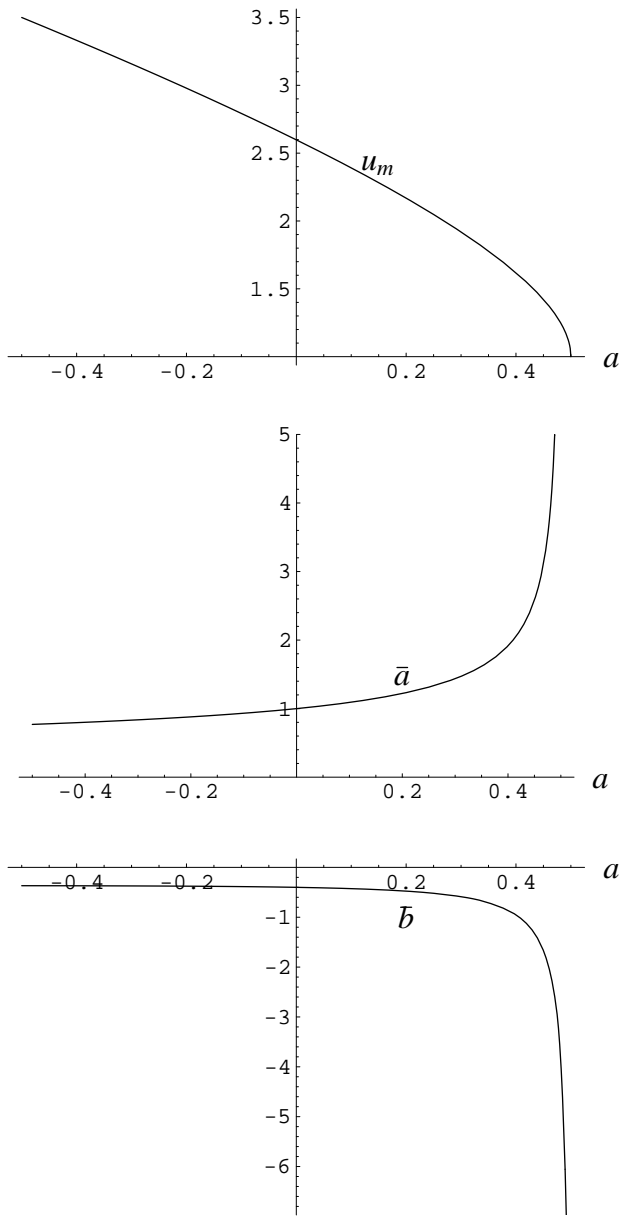}}
 \caption{Coefficients of the strong field limit versus black hole angular momentum.}
 \label{Fig SFL}
\end{figure}

Fig. \ref{Fig SFL} shows the strong field limit coefficients as
functions of $a$. The minimum impact parameter $u_m$ decreases with $a$ in a
way similar to $x_m$. $\overline{a}$ grows
and $\overline{b}$ decreases, both diverging with opposite signs at $a=1/2$.
The divergence of the coefficients of the expansion warns that the
strong field limit deflection angle (\ref{S F L theta}) no longer represents a
 reliable description in the regime of high $a$, since $x_m$ becomes a
higher order singularity in the function $f(z,x_0)$ and a different
expansion should be performed.

\section{Precession of the orbits at small declinations}
\label{Sec pre}

The study of the deflection angle for photons lying on the
equatorial plane is sufficient to write a one-dimensional lens
equation. However, to address the caustic structure and calculate
the magnification of the images, we need a two-dimensional lens
equation. For this reason, in this section we analyze trajectories
close to the equatorial plane. They are described by one further
coordinate: the polar angle $\vartheta$, or equivalently, the
declination $\psi=\frac{\pi}{2}-\vartheta$. The problem becomes
too involved to be solved in general but we shall give a complete
description for quasi-equatorial motion, preserving also the
simplicity and immediacy of the strong field limit scheme.

In order to remain at small declinations, we restrict to light
rays characterized by a small inward inclination $\psi_0$ and
small height $h$ compared to the projected impact parameter $u$,
with $\psi_0 \sim \frac{h}{u}$. Retaining the first relevant
terms, from Eqs. (\ref{J=u cos psi0})-(\ref{Q}) we get
\begin{eqnarray}
&& J \simeq u \\ && Q \simeq h^2+\overline{u}^2 \psi_0^2 \\ &&
\overline{u} \equiv \sqrt{u^2-a^2}.
\end{eqnarray}

We require the declination $\psi$ to stay small (of the order of
$\psi_0$) during the motion. Dividing Eq. (\ref{dottheta}) by Eq.
(\ref{dphids}), we get a simple evolution equation for $\psi$ as a
function of the azimuth $\phi$
\begin{equation}
\frac{d \psi}{d \phi}=\pm \omega(\phi) \sqrt{\overline{\psi}^2
-\psi^2} \label{Eqpsi}
\end{equation}
with
\begin{eqnarray}
&& \overline{\psi}=\sqrt{\frac{h^2}{\overline{u}^2}+\psi_0^2} \label{psibar} \\%
&& \omega(\phi)=\overline{u}\frac{a^2+x(\phi)(x(\phi)-1)}{\left[
a+u(x(\phi)-1) \right]x(\phi)}.
\end{eqnarray}

In the Schwarzschild case ($a=0$), $\omega\rightarrow 1$ and Eq.
(\ref{Eqpsi}) is immediately solved to
\begin{equation}
\psi(\phi) =\overline{\psi} \cos (\phi+\phi_0).
\end{equation}
After each loop around the black hole, the declination returns to
the initial value. This means that there is no precession of the
orbital plane, as expected for a spherically symmetric black hole.

For non-vanishing angular momenta, $\omega$ is no longer a
constant, since $x$ depends on $\phi$, and the solution of Eq. (\ref{Eqpsi})
is generalized to
\begin{equation}
\psi(\phi)=\overline{\psi} \cos \left( \overline{\phi} +\phi_0 \right)
\label{psi}
\end{equation}
with
\begin{equation}
\overline{\phi}=\int\limits_0^\phi \omega(\phi') d \phi'.
\end{equation}

Since we are interested to gravitational lensing, the photon comes from
infinity and returns to infinity. Therefore, we are interested to the quantity
\begin{equation}
\overline{\phi}_f=\int\limits_0^{\phi_f} \omega(\phi') d \phi'.
\end{equation}
where $\phi_f$ is the total azimuthal shift experienced by the photon in its
whole trajectory, given by Eq. (\ref{phif}).

This integral can be rewritten as
\begin{equation}
\overline{\phi}_f=2\int\limits_{x_0}^\infty \omega(x) \frac{d \phi}{dx} dx=
\int\limits_0^1 R_\omega(z,x_0)f(z,x_0) dz
\end{equation}
where
\begin{equation}
R_\omega(z,x_0)=\omega(x)R(z,x_0)
\end{equation}
with $R(z,x_0)$ and $f(z,x_0)$ given by Eqs. (\ref{R}) and (\ref{f}),
respectively. At this point, the integral can be solved by the same technique
 used for the integral (\ref{I z}) with a very similar result.
This is possible since
$\omega(x)$ adds no singularities and can be englobed in the regular function
$R$.

The final result is
\begin{eqnarray}
&& \overline{\phi}_f=-\hat a \log \left( \frac{\theta D_{OL}}{u_m}-1 \right)+
\hat b \label{S F L phibar}\\
&& \hat{a}= \frac{R_\omega(0,x_m)}{2\sqrt{\beta_m}}=1 \label{hata}\\%
&& \hat{b}=-\pi+\hat b_D+\hat b_R+\hat{a} \log \frac{c x_m^2}{u_m}
\label{hatb}
\end{eqnarray}
where
\begin{eqnarray}
&&\! \!\!\!\!\! \hat b_D=2 \hat{a} \log
\frac{2(1-y_m)}{A'_m x_m}
\\ && \! \!\!\!\!\! \hat b_R=\int\limits_0^1\left[
R_\omega(z,x_m)f(z,x_m)-R_\omega(0,x_m)f_0(z,x_m) \right]dz.
\end{eqnarray}

\begin{figure}
\resizebox{\hsize}{!}{\includegraphics{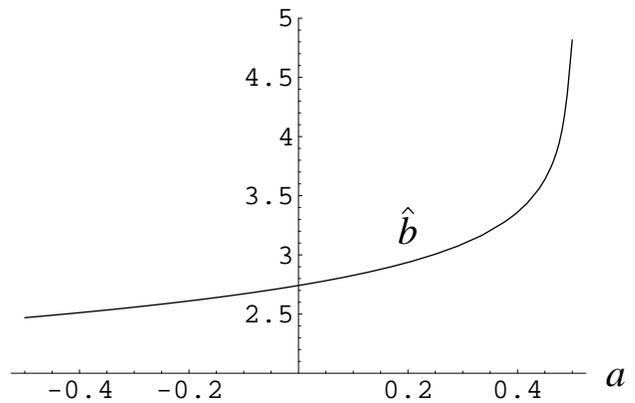}}
 \caption{The coefficient $\hat b$ as a function of $a$. }
 \label{Fig hatb}
\end{figure}

Notice that the coefficient of the logarithmic term $\hat a$ turns ro be
 exactly equal to 1 for
all values of the black hole spin. The coefficient $\hat b$ is just
$\overline{b}+\pi$ in the Schwarzwschild limit $a=0$. This recovers the
equivalence between the phase in the polar motion $\overline{\phi}_f$ and
the total azimuthal shift $\phi_f$ in this limit.
The full behaviour of $\hat b$ is plotted
in Fig. \ref{Fig hatb}. This coefficient diverges as well in the extremal black
hole limit $a \rightarrow 1/2$.

As a general remark, we can say that for positive angular momenta
$\omega$ is always less than one, so that $\overline{\phi}_f<\phi_f$.
As a consequence, the orbital plane
suffers a counterclockwise precession, i.e. after each loop it is
necessary an additional $\Delta \phi$ to reach the same
declination $\psi$. On the contrary, for negative angular momenta,
$\omega>1$. In this case, the
precession is clockwise, i.e. the photon reaches the same
declination before completing a loop.

The integration constant $\phi_0$ in Eq. (\ref{psi}) is fixed by
the initial conditions. In particular, we have to impose that at
$\phi=0$ the declination is just minus the inclination of the
incoming photon trajectory, that we have indicated by $\psi_0$.
The result is that
\begin{equation}
\phi_0=-\mathrm{Sign} [h] \arccos \left[ -
\frac{\psi_0}{\overline{\psi}} \right]. \label{phi0}
\end{equation}

The declination of the outward photon is thus
\begin{equation}
\psi_f \equiv \psi(\phi_f)=\overline{\psi} \cos \left(
\overline{\phi}_f +\phi_0 \right). \label{psif}
\end{equation}

In alternative, using the expression of $\phi_0$, we can write
\begin{equation}
\psi_f =-\psi_0 \cos  \overline{\phi}_f -\frac{h}{\overline{u}}
\sin \overline{\phi}_f. \label{psif2}
\end{equation}

The phase $\overline{\phi}_f$, calculated through Eq. (\ref{S F L phibar}),
 has a central
importance in the discussion of Sects. \ref{Sec pol}, \ref{Sec
Ima}.

\section{Lensing in the equatorial plane} \label{Sec leq}

In the previous sections we have expressed the deflection angle
$\alpha$ as a function of the impact angle $\theta$ and the
outward declination $\psi_f$ as a function of the incoming
inclination $\psi_0$. We are thus ready to write a lens equation
for the Kerr black hole. In this section we shall write the
equatorial lens equation, while the next section will deal with
the polar lens equation.

\begin{figure}
\resizebox{\hsize}{!}{\includegraphics{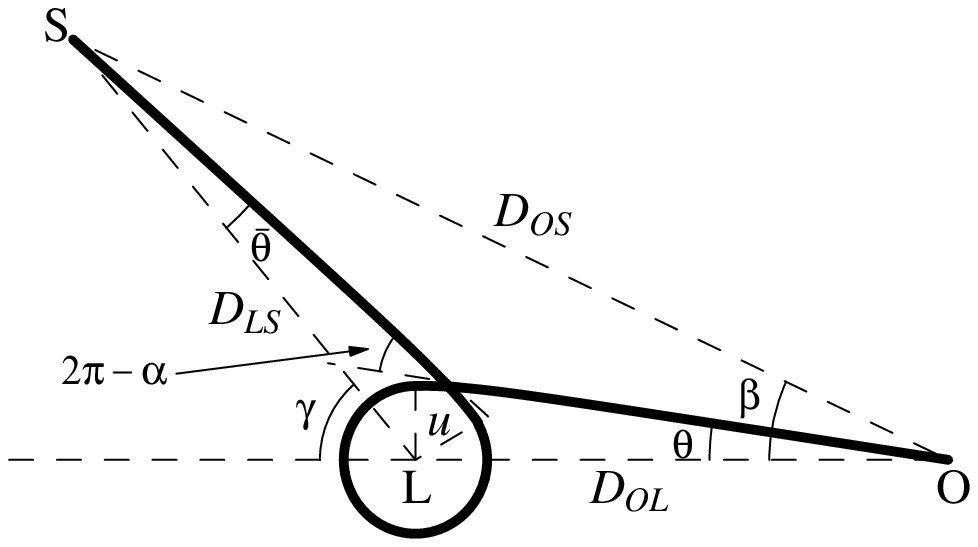}}
 \caption{The lensing geometry projected on the equatorial plane
 in the case of the first relativistic image. $\theta$ is the impact angle
 as seen by the observer, $\beta$ is the angular position of the source
 as seen by the observer, $\gamma$ is the angular position of the source
 as seen by the lens, $\overline{\theta}$ is the impact angle from the source. }
 \label{Fig equ leq}
\end{figure}

So let us start from the ideal case when observer and source both
lie on the equatorial plane of the Kerr black hole and the whole
trajectory of the photon is confined on the same plane. In
previous works, the strong field limit has been developed assuming
an almost perfect alignment of source, lens and observer. This
because, for spherically symmetric metrics, the better the
alignment the higher is the magnification. As we shall see in Sec.
\ref{Sec Ima}, this is no longer the case for Kerr black holes.
Therefore we shall write the equatorial lens equation in a more
general way, allowing for a generic geometric disposition of lens,
source and observer (see Fig. \ref{Fig equ leq}).

The optical axis is the line joining the observer and the lens.
Setting the origin on the black hole, the angle between the
direction of the source and the optical axis will be indicated by
$\gamma$. $\gamma\simeq 0$ is the case of almost perfect alignment
discussed in Refs. \cite{BCIS,ERT,Pet,Boz}. From the lensing
geometry, illustrated in Fig. \ref{Fig equ leq}, we can write the
relation
\begin{equation}
\gamma=-\alpha+ \theta +\overline{\theta} \; \; \; \mathrm{mod} \;
2\pi,
\end{equation}
where
\begin{equation}
\overline{\theta}\simeq \frac{u}{D_{LS}} \simeq
\frac{D_{OL}}{D_{LS}} \theta
\end{equation}
is the impact angle from the source and $D_{LS}$ is the distance
between the lens and the source.

The equatorial lens equation is then
\begin{equation}
\gamma=\frac{D_{OL}+D_{LS}}{D_{LS}}\theta-\alpha(\theta) \; \;
\mathrm{mod} \; 2\pi \label{Lens equation}
\end{equation}

In this lens equation $\gamma$ can assume any value in the
trigonometric interval $[-\pi,\pi]$. The source may even be on the
same side of the observer when $\gamma=\pi$. The relation between
$\gamma$ and $\beta$ (the angular position of the source as seen
by the observer) is
\begin{equation}
\sin \beta= \frac{D_{LS}}{D_{OS}} \sin \gamma,
\end{equation}
but here in general we cannot substitute the sines by their
arguments. $D_{OS}$ is the distance between source and observer
which does not coincide with the distance $D_{OL}+D_{LS}$ actually
covered by lensed photons.

To solve the lens equation, since $\theta = \frac{u}{D_{OL}} \ll
1$, in a first step we solve the Eq. $\gamma=-\alpha(\theta)$ mod
$2\pi$. Using the expression for the deflection angle derived in
the strong field limit (\ref{S F L theta}), we find
\begin{eqnarray}
&& \theta_n^0=\frac{u_m}{D_{OL}} \left(1+e_n \right) \label{theta0}\\%
&& e_n=e^{\frac{\overline{b}+\gamma-2n\pi}{\overline{a}}},
\end{eqnarray}
where $n=1,2,\ldots$ indicates the number of loops done by the
photon around the black hole. This solution is then corrected
expanding $\alpha(\theta)$ around $\theta_n^0$
\begin{eqnarray}
&\alpha(\theta)&=\alpha(\theta_n^0)+ \left. \frac{\partial
\alpha}{\partial \theta} \right|_{\theta_n^0}
(\theta-\theta_n^0)+o(\theta-\theta_n^0) \nonumber \\ && \simeq
-\gamma-\frac{\overline{a} D_{OL}}{u_m e_n}(\theta-\theta_n^0).
\end{eqnarray}
Substituting in (\ref{Lens equation}) and neglecting higher order terms, we find
\begin{equation}
\theta_n\simeq \theta_n^0\left(1-\frac{u_m e_n (D_{OL} + D_{LS})
}{\overline{a} D_{OL}D_{LS}}\right),
\end{equation}
where the correction is much smaller than $\theta_n^0$.

Images are formed on both sides of the lens. As all strong field
limit coefficients depend on $a$ we have to be careful and choose
the correct sign for the angular momentum. Conventionally we call
north the direction of the black hole spin. Then photons winding
counterclockwise are direct and are described by a positive
$a$. They form images on the eastern side of the black hole.
Images formed by retrograde rays appear on the western side and
are described taking a negative $a$ and reversing the sign of
$\gamma$.

\section{Lensing at small declinations} \label{Sec pol}

The lens equation (\ref{Lens equation}) describes trajectories
lying on the equatorial plane and can be employed to calculate the
positions of the relativistic images. However, to investigate the
problem on a deeper level we are forced to study what happens at
least for small displacements from the equatorial plane. In this
section we shall assist Eq. (\ref{Lens equation}) by its polar
counterpart, which is necessary to understand the caustic
structure and compute the magnification of the images.

Thanks to the small declination hypothesis, at the lowest order we
can neglect any backreaction on the equatorial lens equation. In
all our discussion we shall speak (using time reversal) as the
photon were emitted by the observer and absorbed by the source.

\begin{figure}
\resizebox{\hsize}{!}{\includegraphics{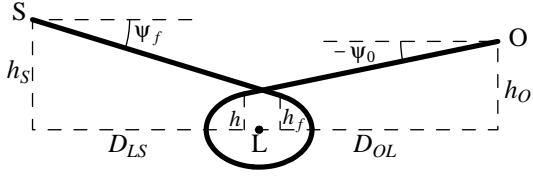}}
 \caption{The lensing geometry projected on the vertical plane. In this picture we have
 assumed $\gamma=0$ for simplicity.}
 \label{Fig Pol leq}
\end{figure}

Consider a source whose height on the equatorial plane is $h_S$.
The height of the observer will be indicated by $h_O$. We shall
assume that the following hierarchy of distances holds (see Fig.
\ref{Fig Pol leq})
\begin{equation}
u \ll (h_O,h_S) \ll (D_{OL},D_{LS}).
\end{equation}

Recalling the meaning of the parameters $\psi_0$ and $h$ used
insofar to identify the incoming light ray, we can write down the
simple geometric relation
\begin{equation}
h=h_O+D_{OL} \psi_0. \label{hO}
\end{equation}

A similar relation holds between the outgoing photon parameters
$h_f$, $\psi_f$ and the source position
\begin{equation}
h_S=h_f+D_{LS} \psi_f. \label{hS}
\end{equation}

Given the positions of source and observer, our
purpose is to determine $\psi_0$, the inclination under which the
observer emits (sees) the light ray.

By symmetry between the outgoing and the incoming parameters, Eq.
(\ref{psibar}) for $\overline{\psi}$ can be written substituting
$\psi_0$ and $h$ by $\psi_f$ and $h_f$
\begin{equation}
\overline{\psi}=\sqrt{\frac{h_f^2}{\overline{u}^2}+\psi_f^2}.
\end{equation}

In this way, we can express $h_f$ in terms of $\psi_f$ and then,
by Eq. (\ref{psif}), in terms of $\overline{\phi}_f$ and $\phi_0$
\begin{equation}
h_f=\overline{u}\overline{\psi} \sin \left( \overline{\phi}_f
+\phi_0\right).
\end{equation}

Recalling Eq. (\ref{phi0}), we also get
\begin{equation}
h_f=-\overline{u} \psi_0 \sin  \overline{\phi}_f -h \cos
\overline{\phi}_f. \label{hf2}
\end{equation}

Substituting in Eq. (\ref{hS}) together with Eq. (\ref{psif2}), we
get
\begin{equation}
h_S=-\psi_0 \overline{u} S-h C -D_{LS} \psi_0
C+D_{LS}\frac{h}{\overline{u}} S,
\end{equation}
where
\begin{eqnarray}
&& S=\sin \overline{\phi}_f \\ && C=\cos \overline{\phi}_f.
\end{eqnarray}

Finally, substituting $h$ from Eq. (\ref{hO}) and discarding
higher order terms, we obtain the lens equation in the polar
direction
\begin{eqnarray}
&h_S=& h_O \left( \frac{D_{LS}}{\overline{u}} S-C \right) -
\nonumber \\ && \psi_0 \left[ (D_{OL}+D_{LS}) C
-\frac{D_{OL}D_{LS}}{\overline{u}} S
  \right]. \label{Polar leq}
\end{eqnarray}

In this equation $\psi_0$ is directly related to the heights of
the observer and the source. The solution is

\begin{equation}
 \psi_{0,n}  = \frac{h_S+ h_O C_n-h_O
\frac{D_{LS}}{\overline{u}} S_n  }{ -(D_{OL}+D_{LS})  C_n
+\frac{D_{OL}D_{LS}}{\overline{u}} S_n }, \label{psi0n}
\end{equation}
where $S_n$ and $C_n$ are $S$ and $C$ calculated for
$\overline{\phi}_f=\overline{\phi}_{f,n}$. The phase
$\overline{\phi}_{f,n}$ of the n-th image is the only quantity
that needs to be calculated preliminarly. However, once the
equatorial lens equation (\ref{Lens equation}) is solved, we know
the impact angle $\theta_n$ of the n-th image and then we can calculate
$\overline{\phi}_f$ by Eq. (\ref{S F L phibar}).

As a consistency check we can see what we obtain in the
Schwarzschild case when the photon completes just one loop around
the black hole, exiting on the opposite side. In this case $a=0$
and $\overline{\phi}_{f,n}=(2n+1)\pi$. We get
\begin{equation}
\psi_{0,n}|_{\overline{\phi_f}=(2n+1)\pi}=\frac{h_S-h_O}{D_{OL}+D_{LS}},
\label{psi0sch}
\end{equation}
which is the correct result for photons passing very close to the
black hole, looping around it.

The consistency of our approximation requires that $\psi_0\ll 1$
and $h\ll u$. From Eq. (\ref{hO}) the height is
\begin{eqnarray}
& h_n & = \frac{h_S D_{OL}- h_O D_{LS} C_n} { -(D_{OL}+D_{LS}) C_n
+\frac{D_{OL}D_{LS}}{\overline{u}} S_n}. \label{hn}
\end{eqnarray}

For a generic $\overline{\phi}_f$, both constraints are
automatically satisfied, since the second term in the denominators
dominates and we have that $\psi_0\sim h_O/D_{OL}$ and $h \sim u
h_O/D_{OL}$. However, in the neighborhood of
$\overline{\phi}_f=k\pi$ the denominators of the two expressions
can vanish, making diverge both quantities. The equation
\begin{equation}
K(\gamma)=\overline{u} (D_{OL}+D_{LS}) C -D_{OL}D_{LS} S=0
\label{Caustic}
\end{equation}
defines the positions of the caustic points. In the next section
we will discuss this equation in connection with the magnification
of the images formed by sources close to the caustic points (which
we call enhanced images for simplicity).

Surprisingly, thanks to the dragging phenomenon, the
quasi-equatorial hypothesis is nearly always satisfied, except for
enhanced images. In this situation the quasi-equatorial motion
hypothesis is satisfied only for particular geometric
configurations which keep $\psi_0$ and $h$ under control.

\section{Magnification and caustic points} \label{Sec Ima}

The magnification is classically defined as the ratio of the
angular area element of the image and the corresponding angular
area element of the source that the observer would see if there
were no lens. The angular area element of the image is
\begin{equation}
d^2A_I= d\theta d\psi_0.
\end{equation}
The distance covered by the photons is $D_{OL}+D_{LS}$ and then
the corresponding angular area element of the source is
\begin{equation}
d^2A_S= \frac{D_{LS} d\gamma dh_S}{(D_{OL} + D_{LS})^2}.
\end{equation}
In fact the source element in the vertical direction is
$dh_S/(D_{OL}+D_{LS})$. In the horizontal direction, the source
element is span by $d\gamma$ when seen from the lens which
corresponds to an angle $D_{LS} d\gamma/(D_{OL} + D_{LS})$ seen
from the observer. If we want to compare the luminosity of a
lensed image with the luminosity of the direct image (namely the
source observed directly along $D_{OS}$ without lensing), the
magnification is to be multiplied by the factor
$\frac{(D_{OL}+D_{LS})^2}{D_{OS}^2}$.

Our lens application has the form
\begin{eqnarray}
&&\gamma=\gamma(\theta) \\ && h_S=h_S(\theta,\psi_0),
\end{eqnarray}
where the dependence on $\theta$ in the polar lens application is
through $\overline{\phi}_f$ and we have neglected the backreaction
of $\psi_0$ on $\gamma$. The ratio between $d\gamma dh_S$ and
$d\theta d\psi_0$ is given by the modulus of the Jacobian
determinant of the lens application
\begin{equation}
|J|=\left| \frac{\partial \gamma}{\partial \theta} \frac{\partial
h_S}{\partial \psi_0}\right|.
\end{equation}

The magnification is then given by
\begin{equation}
\mu=\frac{d^2A_I}{d^2A_S}=\frac{(D_{OL} + D_{LS})^2}{D_{LS}}
\frac{1}{|J|}.
\end{equation}

By the equatorial lens equation (\ref{Lens equation}), retaining
the dominant terms, we have
\begin{equation}
\frac{\partial \gamma}{\partial \theta} \simeq -\frac{\overline{a}
D_{OL}}{u_m e_\gamma},
\end{equation}
with
\begin{equation}
e_\gamma=e^\frac{\overline{b}+\gamma}{\overline{a}}.
\end{equation}

In the following, it is convenient to encode the number of loops
done by the photon within $\gamma$, in order to write more compact
formulae for all the relativistic images. So $\gamma$ can assume
any negative real value; $\gamma$ mod $2\pi$ represents the
angular position of the source and $n=\left[
\frac{\pi-\gamma}{2\pi} \right]$ is the number of loops done by
the photon. In concrete, two values of $\gamma$ differing by a multiple of
$2\pi$ represent the same source position with respect to the
lens, but reached by photons performing a different number of
loops around the lens. For example, $\gamma=0$ is a source aligned
behind the lens reached by a photon making no loop (weak field
lensing); $\gamma=-2\pi$ is the same source behind the lens but
reached by a photon making one loop; $\gamma=-4\pi$ is the same
source for a photon making two loops, and so on.

By the polar lens equation (\ref{Polar leq}), we have
\begin{equation}
 \frac{\partial h_S}{\partial \psi_0}=
(D_{OL}+D_{LS}) C -\frac{D_{OL}D_{LS}}{\overline{u}} S.
\end{equation}

Assembling everything together, we get
\begin{equation}
\mu=\frac{(D_{OL}+D_{LS})^2}{D_{OL}D_{LS}} \frac{\overline{u} u_m
e_\gamma}{\overline{a} \left| \overline{u} (D_{OL}+D_{LS}) C
-D_{OL}D_{LS} S \right|}. \label{mu}
\end{equation}

For a generic $\overline{\phi}_{f}$, $\mu =O\left(
\frac{u}{D_{OL}}\right)^2$, but for the enhanced images, $\mu$ may
even diverge (formally for point-like sources) when the
denominator of Eq. (\ref{mu}) vanishes. The $\gamma$'s where this
happens are called caustic points. At the lowest order in
$\frac{u}{D_{OL}}$, Eq. (\ref{Caustic}) reduces to
\begin{equation}
\overline{\phi}_f \simeq k \pi.
\end{equation}
Combining the formula (\ref{S F L phibar}) for the phase $\overline{\phi}_f$
with the formula for the deflection angle (\ref{S F L theta}) and using the
equatorial lens equation at the lowest order $\gamma=-\alpha(\theta)$,
this equation becomes
\begin{equation}
-\frac{\gamma+\overline{b}}{\overline{a}}+\hat b=k\pi.
\end{equation}

The solutions of this equation determine the angular positions
$\gamma_k$ of the caustic points
\begin{equation}
\gamma_k=-\overline{b}+\overline{a}(\hat{b}-k\pi).
\end{equation}

For each $k$, we have one caustic point for direct photons and one caustic
point for retrograde photons. $k=1$ would
describe the weak field caustic points, formed when the azimuthal
shift is about $\pi$. To be coherent with our strong
field limit approximation, we shall restrict our analysis to $k\geq 2$.

Expanding the denominator of Eq. (\ref{mu}) around the caustic points,
we have
\begin{equation}
K(\gamma)\simeq K'(\gamma_k) (\gamma-\gamma_k(a))= -\frac{D_{OL} D_{LS}}
{\overline{a}}
 (\gamma-\gamma_k(a)). \label{kappa}
\end{equation}

To understand the nature of these caustic points, notice that in
the Schwarzschild limit $\gamma_k \rightarrow -(k-1)\pi$ and
 all the odd caustic points are
aligned on the optical axis behind the lens on consecutive Riemann
folds while the even ones are aligned before the lens. If the source
is aligned behind the lens, the $n$-th image is given by photons
doing $n$ loops around the black hole. Setting $\gamma \sim -2n
\pi$, the closest caustic point is $\gamma_{2n+1}$. Then we
can recover the Schwarzschild magnification for the images created by
a source behind the lens \cite{BCIS}
\begin{eqnarray}
&\mu_n^{Sch}&=\frac{(D_{OL}+D_{LS})^2}{D_{OL}D_{LS}} \frac{u u_m
e_n}{\overline{a} \left| \frac{D_{OL}D_{LS}}{\overline{a}}
 (2n\pi+\gamma) \right|}=
\nonumber
\\ &&=\frac{D_{OS}}{D_{OL}^2 D_{LS}} \frac{u_m^2
e_n(1+e_n)}{\left|\beta \right|}.
\end{eqnarray}

\begin{figure}
\resizebox{\hsize}{!}{\includegraphics{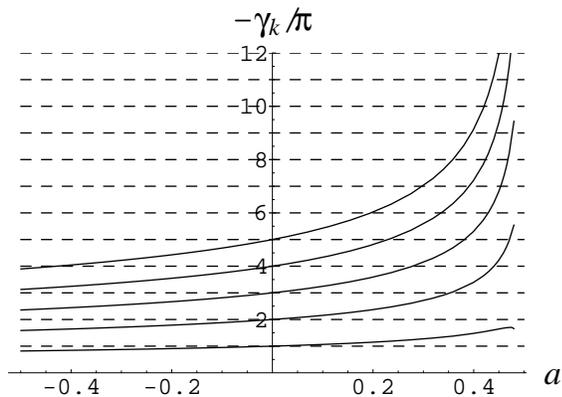}}
 \caption{The angular positions of the first five relativistic
caustic points:
 $k=2,3,4,5,6$ from below to above. When $\gamma=2m\pi$ the source
 is behind the lens, when $\gamma=(2m+1)\pi$ the source is before the
 lens.
 }
 \label{Fig cau}
\end{figure}

In Fig. \ref{Fig cau} we plot the positions of the first five
relativistic caustic points as functions of the black hole angular
momentum. The first relativistic caustic point $\gamma_2$ is
obtained when the photon turns around the black hole and comes
back towards the observer. $\gamma_2$ is thus close to $-\pi$ but
is anticipated for negative $a$ and delayed for positive $a$.
$\gamma_3$ is behind the lens but, at large angular momenta, can
move very far from the initial position. At high values of the
spin, the caustic points drift so much that they can even change
their Riemann fold.

We can specify the magnification formula for the enhanced images
using Eq. (\ref{kappa})
\begin{eqnarray}
&&\mu_k^{enh}=\frac{(D_{OL}+D_{LS})^2}{D_{OL}^2 D_{LS}^2}
\frac{\overline{\mu}_k(a)}{|\gamma -\gamma_k|} \\ &&
\overline{\mu}_k(a)= \overline{u}(\gamma_k(a)) u_m(a)
e_{\gamma_k(a)}.
\end{eqnarray}

The quantity $\overline{\mu}_k$ regulates the magnification close
to caustic points. The dependence on $\gamma$ has been extracted
and has the typical $|\gamma-\gamma_k|^{-1}$ behaviour. The
dependence on the astronomical distances $D_{OL}$, $D_{LS}$,
$D_{OS}$ is negligible in $\overline{\mu}_k$ at the lowest order
in $\frac{u}{D_{OL}}$. So we can use $\overline{\mu}_k$ as a
measure of the magnifying power of the black hole for different
enhanced images and different angular momenta.

\begin{figure}
\resizebox{\hsize}{!}{\includegraphics{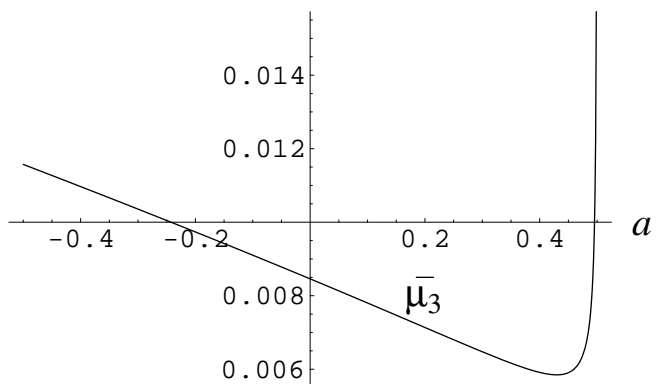}}
 \caption{The magnifying power at the caustic point $\gamma_3$.
 }
 \label{Fig mu3}
\end{figure}

In Fig. \ref{Fig mu3} we plot the magnifying power
$\overline{\mu}_3$ of the caustic point $\gamma_3$, which for
$a=0$ generates the first relativistic image of a source behind
the lens. The magnification grows for negative angular momenta
while decreases for positive $a$ almost linearly. The divergence in the
magnification when $a$ approaches its extremal value $1/2$ should not be taken
seriously, as the standard strong field limit approximation breaks down as
explained in Sect. \ref{Sec equ}.

The shape of $\overline{\mu}_k$ remains more or less the same for
every $k$ but, since $e_{\gamma_k}=e^{\hat{b}-k\pi}$, we have that
\begin{equation}
\frac{\mu_{k+1}}{\mu_k}\simeq e^{-\pi} =0.043.
\end{equation}
The magnification of enhanced images falls quite rapidly as we let
the photons make more and more loops.

\section{Critical curves and caustic structure}

It is well known that the Jacobian of the Schwarzschild lens has
an infinite series of Einstein rings \cite{Nem,VirEll}. The first
one is the classical weak field Einstein ring whose angular radius
is
\begin{equation}
\theta_E= \sqrt{\frac{2D_{LS}}{D_{OL}D_{OS}}}.
\end{equation}
The corresponding caustic is the point at
$\gamma_1=0$.

At small impact parameters we enter the strong field limit of the
Schwarzschild lensing and the light rays wind around the black
hole. The second Einstein ring is created by photons coming back
towards the observer. The caustic is at $\gamma_2=-\pi$.

Decreasing $u$ further, the light ray completes a loop and we have
the third Einstein ring, whose caustic point is $\gamma_3=-2\pi$
and is superposed on the first caustic point (on the second
Riemann fold).

Summing up, the Schwarzschild lens has a large weak field Einstein
ring and an infinite series of concentric relativistic Einstein
rings, very close to the minimum impact angle $\theta_\infty$. In
the region bounded by the $n$-th ring and the $n+1$-th one, the
sign of the Jacobian is $(-1)^{n+1}$.

%

What changes when we turn on the spin of the black hole? As
regards the first Einstein ring of the weak field limit, it is
distorted and shifted. As a consequence, the caustic point turns
into a finite extension diamond shaped caustic \cite{Ser}.

For the critical curves in the strong field limit, we can
calculate their intersections with the equatorial plane, which are
\begin{eqnarray}
&&\theta_k^{cr}\simeq \theta_k^{0,cr}\left(1-\frac{u_m
e_{\gamma_k} (D_{OL} + D_{LS}) }{\overline{a} D_{OL}D_{LS}}\right)
\\
&& \theta_k^{0,cr}=\frac{u_m}{D_{OL}} \left(1+e_{\gamma_k}
\right).
\end{eqnarray}
They are closer to the optical axis on the positive $a$ side (for
left-winding photons) and farther on the negative $a$ side (i.e.
for right-winding photons). Therefore critical curves are
distorted and shifted towards the negative $a$ side, that is the
western side, if north is the direction of the spin.

\begin{figure}
\resizebox{\hsize}{!}{\includegraphics{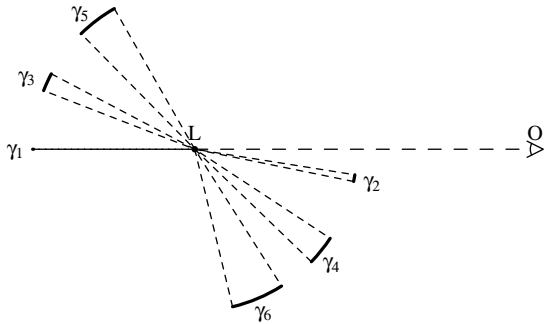}}
 \caption{The first six caustics of the Kerr lens for $a=0.1$,
 marked by the thick lines between $\gamma_{k}(-|a|)$
 and $\gamma_k(|a|)$.
 }
 \label{Fig cau2}
\end{figure}

The caustics are no longer points but acquire a non vanishing
extension. $\gamma_k(-|a|)$ and $\gamma_k(|a|)$ represent the
intersections of the $k$-th caustic with the equatorial plane. As
$|\gamma_k(-|a|)|<k\pi$ and $|\gamma_k(|a|)|>k\pi$ the caustic is
shifted towards the western side. To visualize this situation, in
Fig. \ref{Fig cau2}, we have plotted the projections of the
caustics on the equatorial plane, as seen from the north
direction. The $[\gamma_1(-|a|),\gamma_1(|a|)]$ caustic is the
weak field one, which stays almost aligned on the optical axis,
while the relativistic caustics drift in the clockwise direction.
As $k$ grows, the caustics become larger and farther from their
initial position on the optical axis. Examining Fig. \ref{Fig cau}
we notice that at high angular momenta the caustics may become
very large, covering even several Riemann folds!

At the lowest order in $\psi$ and neglecting any backreaction on
the equatorial lens equation it is not possible to give a rigorous
classification of the type of catastrophes we encounter on the
equatorial plane. However, the fact that the first caustic in the
weak field assumes the typical diamond shape of quadrupole lenses
suggests a similar picture for strong field caustics. If this is
the case, then the caustic points $\gamma_k$ on the equatorial
plane are cusps. This is consistent with the fact that if we let
$\gamma$ decrease below some $\gamma_k$ the corresponding image
changes parity. This can happen only when at the critical point
two images are formed with the same parity of the original image
\cite{SchWei,Rhi}. These images rapidly move in the vertical
direction and are missed in our quasi-equatorial approximation.

\section{Phenomenological implications}

After all the analysis of the quasi-equatorial lensing in the Kerr
spacetime, we are able to discuss the phenomenological relevance
of the black hole spin.

In Ref. \cite{Boz}, it was shown that all spherically symmetric
black holes produce the same patterns of relativistic images.
These patterns differ by the separations and the luminosity ratios
between different relativistic images. The conclusion was that if
strong field gravitational lensing will be caught by future VLBI
experiments, it may provide a means to distinguish between
different classes of black holes.

In spherically symmetric black hole lensing, a generic source not
aligned with the optical axis produces extremely faint
relativistic images. On the contrary, a point source perfectly
aligned along the optical axis produces (theoretically) infinitely
bright images. Actually the finite source radius cuts off the real
brightness of the images. The relativistic images are maximally
amplified altogether since all the caustic points of spherically
symmetric black holes lie on the optical axis.

In Kerr lensing, the situation becomes radically different. The
crucial fact is that the caustics no longer lie on the optical
axis but drift throughout the trigonometric interval. Then if the
source is close to one caustic point, it cannot be close to any
other. The consequence is that only one image at a time can be
enhanced, while all the others stay extremely faint.

To clarify the situation, suppose we have a source aligned with
the caustic point $\gamma_3(|a|)$. Then the outermost relativistic
image on the eastern side will be enhanced. If we put the source
on $\gamma_3(-|a|)$, then we only enhance the first relativistic
image on the western side. If we put the source on
$\gamma_5(|a|)$, then the second relativistic image on the eastern
side will be enhanced and so on.

Rather than seeing an infinite series of relativistic images on
each side of the lens, we would observe only one enhanced
relativistic image. It would be difficult to recognize a single
image as a gravitational lensing phenomenon rather than any kind
of environmental source around the black hole. Even if we managed,
it would be quite tricky to extract information about the strong
fields around the black hole from one single relativistic image.

However, if the source is inside the caustic, two additional
images should appear, making easier to recognize their real nature
of lensed images of the same source. Moreover, three images can be
used to investigate the gravitational field around the black hole
and put constraints on its parameters. Alas, the additional images
are missed in our quasi-equatorial approximation. In order to
catch them, it is necessary to face the problem of Kerr lensing in
its general form. A reliable treatment of non equatorial images,
would complete our quasi-equatorial study and would make possible
a detailed investigation of strong field gravitational lensing for
high values of the black hole spin.

By now, we have discussed the two extreme situations: spherically
symmetric black holes ($a=0$) and high spin black holes. It is
interesting to estimate the value of the spin which separates the
two regimes. We shall do it, referring to the black hole at the
center of our Galaxy, assuming $D_{OL}=8.5$ kpc, $D_{LS}=1$ kpc.

In order to consider a black hole as spherically symmetric, the
caustic drift must be negligible when compared to the extension of
the source. In fact, in this case, the source does not ``see''
different caustics but they behave roughly as they were all at the
same point. For small $a$ the $\gamma_k$ scale linearly as $-(k-1)\pi
- a(\eta_0+\eta_1 k) $, with $\eta_0$ and $\eta_1$ numerical factors
 of order one. Then,
the drift between two consecutive caustics will be negligible if
\begin{equation}
|\gamma_k(a)-\gamma_{k+2}(a) \; \; \mathrm{mod} \; \; 2\pi|=2\eta_1
a\ll \frac{R_S}{D_{LS}}.
\end{equation}

If we consider a source with radius $R_S=10R_\odot$, we then find
that $a$ should be lower than $10^{-10}$. When $a$ is greater than
this value, the source will see only one caustic at a time.
However, the caustic will be still seen as point-like, since the
extension of the caustic scales as $ a^2$. Therefore, we will
still have two enhanced images, corresponding to the two
intersections of the $k$-th caustic with the equatorial plane
$\gamma_k(a)$ and $\gamma_k(-a)$. This intermediate situation
takes place as long as
\begin{eqnarray}
&|\gamma_k(a)+(k-1)\pi|&-|\gamma_k(-a)+(k-1)\pi|= \nonumber \\
&& (\xi_0+\xi_1 k) a^2 \ll \frac{R_S}{D_{LS}}
\end{eqnarray}
with $\xi_0$ and $\xi_1$ of order one.

For a $10 R_\odot$ source, this requires $a\ll 10^{-5}$. Beyond
this value, only one image at a time will be enhanced (together
with an eventual additional pair of images if the source is inside
the caustic), while all the others stay invisible.

These estimates reveal that the phenomenology of spherically
symmetric black holes is realistic only for black holes with tiny
spin. Yet, as recalled in the introduction, the first estimates of
the spin of the black hole at the center of our Galaxy push
towards high values \cite{LiuMel}. If these estimates are
confirmed, then we are forced to include spin in any realistic
treatment of strong field gravitational lensing for Sgr A*.

\section{Summary}

In this paper we have explored the modifications to strong field
limit gravitational lensing induced by the rotation of the central
body, analyzing the quasi-equatorial null geodesics.

The most apparent change is the formation of extended caustics
which, for high angular momenta, can cover several Riemann folds.
This situation is radically different from spherically symmetric
black holes where the caustics are points aligned behind and in
front of the lens. While for $a=0$ a source behind the lens is
simultaneously close to all odd caustics and gives rise only to
enhanced images, for Kerr black holes the source can be close to
one caustic at a time and thus produces only one enhanced image.

As secondary interesting effects, we can also mention the
asymmetry between images formed by photons winding in the same
sense of the black hole and photons winding in the opposite sense,
the latter appearing farther from the black hole. The magnification
decreases with the spin, being higher for retrograde images.

The study of quasi-equatorial Kerr gravitational lensing is very
instructive and has allowed us to discover a great number of
interesting features of spinning black holes. However, to address
the phenomenology of the black hole at the center of our Galaxy
and/or other black holes with deeper insight, further
investigation is necessary. In fact, we need a punctual
description of the caustic structure not limited to the equatorial
plane. The existence of extended caustics suggests the formation
of pairs of non-equatorial images, missed in our approximation,
which are of striking importance for the phenomenology.

The quasi-equatorial lensing, studied in this work, then
represents a first fundamental step to understand lensing by
spinning black holes. However, the complexity of the problem
requires a global approach in order to give correct and complete
answers to all observational questions. This settles as the main
objective for future work on strong field gravitational lensing.

\begin{acknowledgments}
I am grateful to Mauro Sereno for useful discussions on the
subject.

\end{acknowledgments}

\end{document}